\documentclass[aps,prl,superscriptaddress,showpacs,twocolumn,nofootinbib,longbibliography,notitlepage]{revtex4-1}
\usepackage[pdftex]{graphicx}
\usepackage{amsmath}
\usepackage{xspace}
\usepackage{color}
\usepackage[colorlinks=true,linkcolor=blue,urlcolor=blue,citecolor=blue]{hyperref}
\usepackage[normalem]{ulem}
\usepackage{mathtools}
\usepackage{bm,bbold,amsfonts,mathrsfs,amssymb,dsfont}

\usepackage{enumitem}

\usepackage{textcomp}

\usepackage{hyperref}
\newcounter{myequation}
\makeatletter
\@addtoreset{equation}{myequation}
\makeatother

\newcounter{myfigure}
\makeatletter
\@addtoreset{figure}{myfigure}
\makeatother

\DeclareMathOperator{\tr}{\mbox{tr}}
\DeclareMathOperator{\re}{\mbox{Re}}
\DeclareMathOperator{\im}{\mbox{Im}}

\newcommand{\U}{{\mathcal U}}

\renewcommand{\L}{{\mathcal L}}

\renewcommand{\Re}{{\mathrm{Re}}\,}

\newcommand{\nmax}{n_{\rm max}}
\newcommand{\lmax}{l_{\rm max}}

\usepackage{color}
\definecolor{BV}{rgb}{0.1,0.,0.6}
\definecolor{R}{rgb}{0.9,0,0}
\definecolor{G}{rgb}{0.2,0.8,0.2}
\usepackage{comment}

\newcommand{\ii}{{\mathbf i}}
\newcommand{\jj}{{\mathbf j}}

\renewcommand{\L}{{\mathcal L}}

\newcommand{\be}{\begin{equation}}
\newcommand{\ee}{\end{equation}}

\begin{document}

	\title{
Pseudomode expansion of many-body correlation functions
 }

	\author{ Alexander Teretenkov}
\affiliation{%
		Skolkovo Institute of Science and Technology\\
		Bolshoy Boulevard 30, bld. 1, Moscow 121205, Russia
	}
\affiliation{Department of Mathematical Methods for Quantum Technologies,\\
		Steklov Mathematical Institute of Russian Academy of Sciences\\
		8 Gubkina St., Moscow 119991, Russia
	}

	\author{Filipp Uskov}
	\affiliation{%
		Skolkovo Institute of Science and Technology\\
		Bolshoy Boulevard 30, bld. 1, Moscow 121205, Russia
	}

		%
	\author{Oleg Lychkovskiy}
%
%
	\affiliation{%
		Skolkovo Institute of Science and Technology\\
		Bolshoy Boulevard 30, bld. 1, Moscow 121205, Russia
	}%
\affiliation{Department of Mathematical Methods for Quantum Technologies,\\
		Steklov Mathematical Institute of Russian Academy of Sciences\\
		8 Gubkina St., Moscow 119991, Russia
	}

	\date{\today}








\date{\today}

\begin{abstract}
We present an expansion of a many-body correlation function in a sum of pseudomodes -- exponents with complex frequencies that encompass both decay and oscillations.  The pseudomode expansion emerges in the framework of the Heisenberg version of the recursion method. This method  essentially solves Heisenberg equations in a Lanczos tridiagonal basis constructed in the Krylov space of a given observable. To obtain pseudomodes, we first add artificial dissipation satisfying the dissipative generalization of the universal operator growth hypothesis, and then take the limit of the vanishing dissipation strength. Fast convergence of the pseudomode expansion is facilitated by the localization in the Krylov space, which is generic in the presence of dissipation and can survive the limit of the vanishing dissipation strength.
As an illustration, we present pseudomode expansions of infinite-temperature autocorrelation functions in the quantum Ising and $XX$ spin-$1/2$ models on the square lattice. It turns out  that it is enough to take a few first pseudomodes to obtain a good approximation to the correlation function.
\end{abstract}

	\pacs{}

\maketitle

\medskip
\noindent{\it Introduction.}~ Expanding a function into a functional series is a basic technique with countless applications in mathematics and theoretical physics. Typically, the convergence of the expansion is better (in terms of the domain and the rate) when the terms of the series share some general features of the function being expanded. For example, Fourier series are better suited for periodic functions,  Taylor series --  for polynomially growing functions {\it etc}. Choosing a suitable series  can be decisive when an unknown function is found step by step by computing consecutive terms of the series.

Here we report an expansion technique for many-body time-dependent correlation functions. Generically correlation functions decay exponentially at large times, often with oscillations on top of the exponential decay \cite{Mori_1965_Continued-fraction,Prosen_2002_Ruelle,Fine_2004_Long-time, Fine_2005_Long-time,Morgan_2008_Universal,Sorte_2011_Long-time, Meier_2012_Eigenmodes,Barocchi_2012_Exact,Barocchi_2014_Exponential,Heveling_2022_Stability}.  It was therefore proposed to expand correlation functions in decaying modes $\Re e^{\lambda_l t}$, where $\lambda_l$, $l=0,1,\dots$ are  complex numbers with negative real parts \cite{Mori_1965_Continued-fraction,Prosen_2002_Ruelle,Fine_2004_Long-time, Fine_2005_Long-time,Prosen_2007_Chaos, Morgan_2008_Universal,Sorte_2011_Long-time, Meier_2012_Eigenmodes,Barocchi_2012_Exact,Barocchi_2014_Exponential,Mori_2024_Liouvillian-gap}.

Analogous expansions recurrently appeared in the studies of open systems interacting with large reservoirs \cite{garraway1996cavity, garraway1997nonperturbative, dalton2001theory,Chin_2010_Exact,tamascelli2018nonperturbative, teretenkov2019pseudomode, tamascelli2019efficient, pleasance2020generalized, kanazawa2024standard,Teretenkov_2024_Exact}. Modes with complex frequencies were referred to as {\it  pseudomodes} in some of these studies.  We adopt this term, even though there are no explicit reservoirs in our setting.

Within  a parallel line of research,  it was argued \cite{Prosen_2002_Ruelle,Prosen_2007_Chaos,Sorte_2011_Long-time} that complex modes emerge without any explicit reservoir as quantum analogs of classical Ruelle-Pollicott resonances \cite{Ruelle_1986_Resonances,Pollicott_1985_Rate}.
Recently, this research direction has been revigorated \cite{Mori_2024_Liouvillian-gap}.

From yet another perspective, the expansion of autocorrelation function in terms of $e^{\lambda_l t}$ can be viewed as a consequence of the expansion of the observable in the basis of {\it dynamical symmetries}, which are eigenoperators of the evolution superoperator, with the corresponding eigenvalues $\lambda_l$ admitting, in general, complex values for systems in the thermodynamic limit \cite{Buca_2023_Unified}.

We present a concrete scheme to construct the pseudomode expansion based on (the Heisenberg version of) the recursion method \cite{Mori_1965_Continued-fraction,Dupuis_1967_Moment,viswanath2008recursion,Nandy_2025_Quantum}. This method amounts to constructing a Lanczos basis in the Krylov space of operators and solving coupled Heisenberg equations in this basis. In the many-body case one additionally has to extrapolate the sequence of Lanczos coefficients according to the universal operator growth hypothesis (UOGH)~\cite{Parker_2019}. Recent years have witnessed a considerable progress in applying the recursion method to compute correlation functions in one \cite{Parker_2019,Khait_2016_Spin,deSouza_2020_Dynamics,Yates_2020_Lifetime,Yates_2020_Dynamics,Yuan_2021_Spin,Wang_2024_Diffusion,Bartsch_2024_Estimation,Uskov_Lychkovskiy_2024_Quantum} and two \cite{Uskov_Lychkovskiy_2024_Quantum,Bhattacharyya_2024_Metallic} spatial dimensions.


To obtain the pseudomode expansion within the recursion method, we first add an artificial dissipation to the evolution superoperator, then diagonalize it and finally take the limit of vanishing dissipation strength. In fact, the usage of an auxiliary infinitesimal dissipation is ubiquitous in theoretical physics  -- in particular, it  recurrently appears in various perturbative expansions and diagrammatic techniques \cite{mattuck1992guide,zubarev1996statistical}. Variations of this trick have been used to track nonperturbative Heisenberg evolution \cite{Prosen_2002_Ruelle,Prosen_2007_Chaos,Rakovszky_2022_Dissipation-assisted,White_2023_Effective,Yi-Thomas_2024_Comparing,Kuo_2024_Energy,
Mori_2024_Liouvillian-gap,ermakov2024unified,srivatsa2024probing}.

An important distinctive feature of our approach is the way we add the auxiliary dissipation. We do so directly in the Lanczos basis, and ensure that  the dissipative term obeys the dissipative version of the UOGH \cite{bhattacharya2022operator,Liu_2023_Krylov,Bhattacharjee_2023_Operator}. Such a dissipative term leads to the localization in the Krylov space \cite{Liu_2023_Krylov} (see also \cite{Bhattacharjee_2023_Operator}, \cite{Teretenkov_2024_Exact}). In turn, the localization ensures the emergence of discrete pseudomodes and the fast convergence of the pseudomode expansion.

The rest of the paper is organized as follows. First we introduce the general formalism of the recursion method. After that we present the technique to compute pseudomode eigenvalues and amplitudes within the recursion method. Then we illustrate all major features of the technique, including localization and fast convergence, in a toy exactly solvable model of evolution superoperator. Finally, we explain how to practically construct the truncated pseudomode expansion starting from a many-body Hamiltonian, and illustrate the method by applying it to the quantum Ising and $XX$ spin-$1/2$ models on the square lattice.

\medskip
\noindent{\it Recursion method.}~ Consider some observable given by a self-adjoint Shr\"odinger operator $A$. The same observable in the Heisenberg representation reads $A_t=e^{i t H} A \, e^{-i t H}$, where $H$ is the Hamiltonian of the system. The Heisenberg operator $A_t$ evolves within the {\it Krylov space} spanned by Shr\"odinger  operators $A,\L A,\L^2 A,\dots$, where $\L\equiv [H,\bullet]$ is the evolution {\it superoperator}. Knowledge of $A_t$ allows one to compute time-dependent correlation functions related to this observable. Throughout the paper we focus on the simplest case of infinite-temperature autocorrelation function
\begin{equation}\label{autocorrelation function}
C(t)\equiv \tr\big(  A\,A_t \big)/\tr A^2.
\end{equation}
It has the properties $C(0)=1$ and $C(-t)=C(t)$.

We introduce a scalar product in the space of operators according to $\big(A|B\big)\equiv \tr \big(A^\dagger B\big)/d$,
where $d$ is the Hilbert space dimension (which is assumed to be finite). The scalar product entails the norm $\|A\|=\sqrt{(A|A)}$.  In this notation, the autocorrelation function can be written as $C(t)=(A|A_t)/\|A\|^2$. The superoperator $\L$ is self-adjoint with respect to this scalar product.

The Heisenberg equation of motion reads $\partial_t A_t=i\, \L A_t$. In the recursion method \cite{Mori_1965_Continued-fraction,Dupuis_1967_Moment,viswanath2008recursion,Nandy_2025_Quantum}, this equation is addressed in the orthonormal Lanczos basis $\{O^0,O^1,O^2,\dots\}$ constructed as follows: $O^0=A/\|A\|$, $A^1=\L O^0$, $b_1=\|A^1\|$, $O^1=b_1^{-1}\,A^1$ and
$A^n={\cal L}\, O^{n-1}- \,b_{n-1} O^{n-2}$, $b_n=\|A^n\|$, $O^n=b_n^{-1} A^n$ for $n\geq2$. The coefficients $b_n$ are known as {\it Lanczos coefficients}.    In this basis the evolution superoperator has a tridiagonal form,
\begin{equation}\label{L}
\L=
\begin{pmatrix}
    a_0  & b_1 & 0  & 0  & \dots  \\[0.3 em]
   b_1 & a_1  & b_2 & 0  & \dots \\[0.3 em]
     0 & b_2  & a_2 &  b_3  &  \dots \\[0.3 em]
     0 & 0  & b_3 & a_3  &  \dots \\[0.3 em]
\vdots & \vdots & \vdots & \vdots & \ddots
  \end{pmatrix},
\end{equation}
where, for the moment, $a_n=0$. The coupled Heisenberg equation can be conveniently written in  matrix form as
\begin{align}\label{Heisenberg equations}
\partial_t O_t= & i\,\L O_t,\qquad O_0=O,
\end{align}
where $O_t=\|O^n_t\|_{n=0,1,2,\dots}$ is a column vector constructed of Heisenberg operators $O^n_t$ and $O=\|O^n\|_{n=0,1,2,\dots}$ is a column vector constructed of Schr\"odinger operators $O^n$.

In practice, one can explicitly compute only a finite number $\nmax$ of Lanczos coefficients. The rest of the coefficients should be extrapolated. 
The extrapolation is facilitated by the universal operator growth hypothesis~\cite{Parker_2019} (see also a precursor work in \cite{Liu_1990_Infinite-temperature,Zobov_2006_Second,Elsayed_2014_Signatures}) which states  that in a generic quantum many-body system $b_n$ scales linearly with $n$ at large $n$ (with a logarithmic correction for one-dimensional systems). This asymptotic behavior has been confirmed in various many-body models \cite{Parker_2019,Noh_2021,Heveling_2022_Numerically,Wang_2024_Diffusion,Uskov_Lychkovskiy_2024_Quantum,De_2024_Stochastic,Loizeau_2024_Quantum}. For our approach to be successful, one needs  to compute a  number of Lanczos coefficients large enough that the asymptotic behavior predicted by the UOGH can be identified. The latter requirement can be fulfilled with the help of the state-of-the-art computer codes \cite{Parker_2019,Noh_2021,Heveling_2022_Numerically,Wang_2024_Diffusion,Uskov_Lychkovskiy_2024_Quantum,De_2024_Stochastic,Loizeau_2024_Quantum}, with a clear promise for further improvement.


\begin{figure}[t] 
		\centering
		\includegraphics[width=\linewidth]{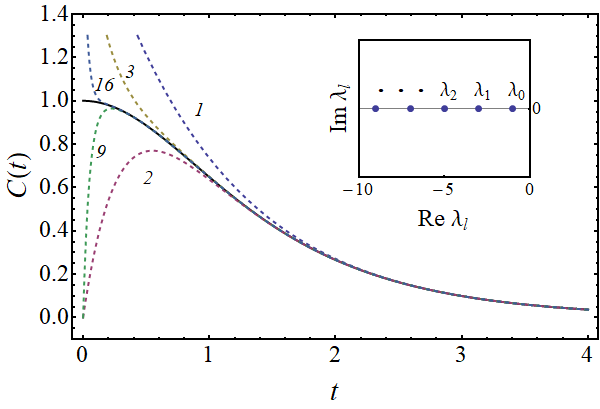}
		\caption{Pseudomode expansion for the toy model. Solid line -- exact autocorrelation function \eqref{C(t) toy model}. Dashed lines -- the pseudomode expansion \eqref{pseudomode expansion toy model} truncated at various orders $\lmax$ (indicated by numbers). Inset - the pseudomode spectrum of the model.  }
		\label{fig_toy_model}
	\end{figure}

\medskip
\noindent{\it Pseudomode expansion.}~ A well-known strategy to improve an approximate solution of a dynamical equation is to add an artificial dissipation and then to consider the limit of the vanishing dissipation strength \cite{Prosen_2002_Ruelle,Prosen_2007_Chaos,Rakovszky_2022_Dissipation-assisted,White_2023_Effective,Yi-Thomas_2024_Comparing,Kuo_2024_Energy,
Mori_2024_Liouvillian-gap}. We implement this general idea by introducing diagonal terms
\begin{equation}\label{a_n}
    a_n=
\begin{cases}
 0,& n \leq n_{\rm max}\\
 i\gamma(n- n_{\rm max}),              & n\geq n_{\rm max}+1
\end{cases}
\end{equation}
in the evolution superoperator \eqref{L}, where $\gamma>0$ is the dissipation strength.

There are good reasons to choose this particular form of artificial dissipation. First of all, any self-adjoint superoperator of Markovian dissipative evolution can be brought to the tridiagonal form \eqref{L} \cite{bhattacharya2022operator,Bhattacharjee_2023_Operator}. Further,  first $(\nmax+1)$ derivatives of $C(t)$ at $t=0$ remain unaltered by the artificial dissipation \eqref{a_n}, thanks to zero diagonal terms. Therefore the artificial dissipation does not alter the initial behavior of the correlation function which is known to be reproduced extremely accurately by the recursion method \cite{Uskov_Lychkovskiy_2024_Quantum}.

Finally and most importantly, the linear growth of $a_n$ at large $n$ conforms with the dissipative version of the UOGH that is believed to be valid for generic Markovian dissipative systems \cite{bhattacharya2022operator,Liu_2023_Krylov,Bhattacharjee_2023_Operator}. A remarkable  feature of such systems is the localization of eigenvectors of $\L$ in the Krylov space \cite{Liu_2023_Krylov} (see also \cite{Bhattacharjee_2023_Operator}, \cite{Teretenkov_2024_Exact}).\footnote{A different mechanism for Krylov  localization was proposed in ref. \cite{Rabinovici_2022_Krylov} (see also \cite{Yates_2020_Lifetime,Yates_2020_Dynamics}) -- an Anderson-type localization in the absence of dissipation due to the disorder in the subleading terms of the Lanczos coefficients. Our experience with specific  systems \cite{Uskov_Lychkovskiy_2024_Quantum} suggests that while such a disorder is indeed present, it may not be strong enough to cause  localization for generic systems and observables.}

Let us discuss the above localization in more detail. Generically, $\L$ is diagonalizable,
\begin{equation}\label{L in diagonal form}
i\,\L \, \U =\U \, \Lambda,
\end{equation}
where $\U$ is an invertible matrix with column vectors being  eigenvectors of  $\L$, and $\Lambda={\rm diag} (\lambda_0,\lambda_1,\lambda_2,\dots)$ is the diagonal matrix of the corresponding eigenvalues. Since $\L=\L^\mathsf{T}$, one can choose $\U$ such that $\U^{-1}=\U^\mathsf{T}$, which is equivalent to normalizing $\U$ according to $\U^\mathsf{T}\U=\mathds{1}$ \cite{Teretenkov_2024_Exact}. We employ such normalization in what follows.

\begin{figure*}[t] 
		\centering
\includegraphics[width=\linewidth]{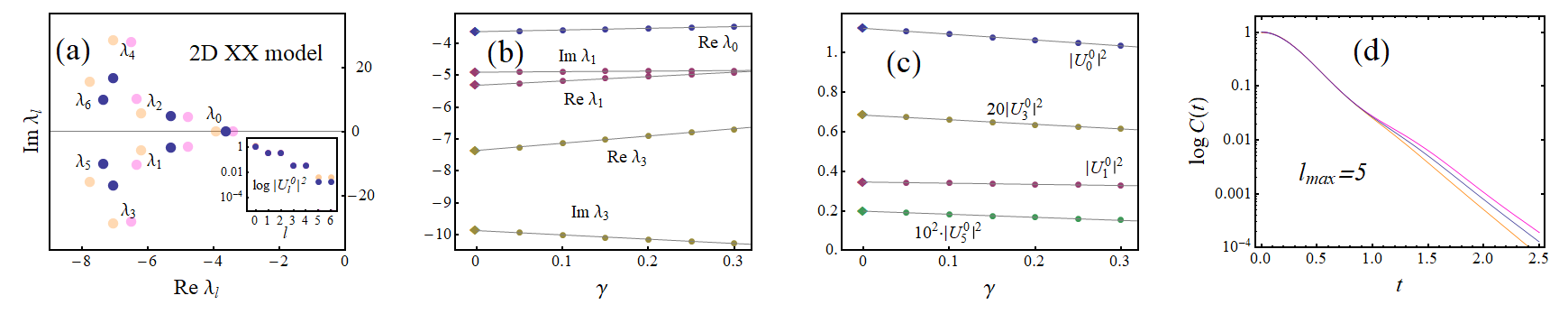}\\
\includegraphics[width=\linewidth]{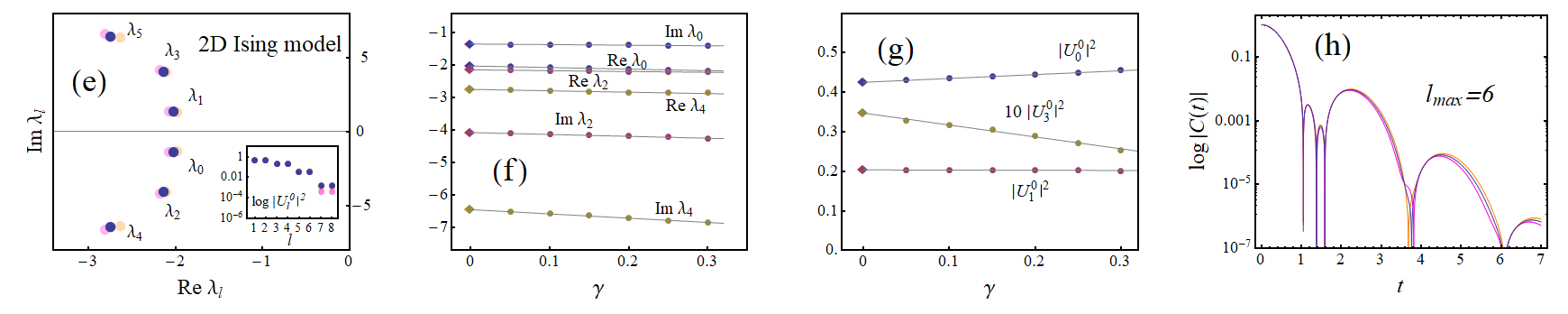}
		\caption{Pseudomode spectra and expansions for two quantum spin-$1/2$ models on a square lattice:  upper row -- $XX$ model~\eqref{H 2D XX}, lower row -- Ising model~\eqref{H 2D Ising}. {\bf (a),(e)}: The rightmost part of the pseudomode spectrum; the inset shows the pseudomode amplitudes. Points of different colors correspond to different ways to extrapolate the Lanczos coefficients; their spread reflects the uncertainty due to the extrapolation error (see Supplemental Material \cite{supp_pseudomode} for details). {\bf(b),(f)} and {\bf(c),(g)}: The convergence of pseudomode eigenvalues and amplitudes, respectively, in the limit of vanishing dissipation strength $\gamma$.  Points indicate the values obtained for finite $\gamma$ in the numerical thermodynamic limit (see Supplemental Material \cite{supp_pseudomode} for details).  Diamonds correspond to the limit of $\gamma\rightarrow 0$ obtained by extrapolation (lines).  {\bf(d),(h)}: The autocorrelation function $C(t)$ computed by means of the truncated pseudomode expansion with  $l_{\rm max}$ pseudomodes. Different lines indicate the effect of the error in Lanczos coefficient  extrapolation, with the color code consistent with (a),(e).  Note that in the Ising model  $C(t)$ has alternating sign, for this reason it is the absolute value of  $C(t)$ which is plotted in (h).   }
		\label{fig spin models}
\end{figure*}

The eigenvalues are either real or come in complex-conjugate pairs. In any case  $\re \lambda_l\leq0$.  We order the eigenvalues by the ascending absolute values of their real parts (complex-conjugate pairs $(\lambda_l,\lambda_{l+1}=\lambda_l^*)$ are additionally ordered according to $\im \lambda_l<0$ ), see Figs.~\ref{fig_toy_model},\ref{fig spin models}.

The localization implies that, for any finite $\gamma$,
\begin{itemize}
\item the spectrum $\{\lambda_0,\lambda_1,\lambda_2,\dots\}$ of $\L$ is discrete and
\item the eigenvectors of $\L$ are localized, $\big|\U_l^n\big|<e^{-\kappa_l |n-l| }$, where $\kappa_l>0$ is the inverse localization length.
\end{itemize}

Our central set of assumptions that underlie the pseudomode expansion concerns the behavior of $\lambda_l$ and $\U_l^0$ in the  limit $\gamma\rightarrow +0$ of the vanishing dissipation strength. Namely, we assume that
\begin{itemize}
\item $\lim\limits_{\gamma\rightarrow +0} \U_l^0$ exists for all $l$, is bounded in absolute value by a constant independent of $l$ and is nonzero at least for some $l$;
\item $\lim\limits_{\gamma\rightarrow +0} \lambda_l$ exists  and has a strictly negative real part for all $l$\footnote{In general, the latter requirement can  be relaxed for the leading pseudomode $\lambda_0$, which can converge to zero. This would imply that $C(t)$ converges to a nonzero value in the limit $t\rightarrow \infty$. We do not address this case in the present paper, leaving it for further studies.} (this statement has been recently rigorously proven for a certain random circuit \cite{Yoshimura_2024_Robustness});
\item the spectrum remains discrete, i.e.
\begin{align}
\lim\limits_{\gamma\rightarrow +0}(\lambda_l-\lambda_{l'})& \neq 0\quad{\rm for} \quad l\neq l'.\label{assumption 2}
\end{align}
\end{itemize}
The above assumptions will be verified  in a toy model of $\L$ and in two microscopic spin models, see Figs.~\ref{fig_toy_model},\ref{fig spin models}. The two assumptions concerning $\lambda_l$ essentially coincide with those underlying quantum many-body  Ruelle-Pollicott
resonances  \cite{Prosen_2002_Ruelle,Prosen_2007_Chaos,Mori_2024_Liouvillian-gap}.
Note that while localization of $\lim\limits_{\gamma\rightarrow +0} \U_l^0$ is not strictly necessary for the pseudomode expansion, as illustrated below in a toy model, such localization is expected generically and is indeed found in microscopic models.

After a standard algebra, one obtains
$
O_t= \U \, e^{\Lambda t} \, \U^\mathsf{T} O
$
and hence
\begin{equation}\label{pseudomode expansion}
C(t)=\sum_{l=0}^\infty \big( \U_l^0 \big)^2 \,e^{\lambda_l t}.
\end{equation}
In this formula, $\U_l^0 $ and $\lambda_l$ should be understood as limiting values at $\gamma\rightarrow +0$. We refer to eq.\eqref{pseudomode expansion} as the pseudomode expansion. Crucially,  the pseudomode expansion is a discrete sum, thanks to the assumption \eqref{assumption 2}.

To the best of our knowledge, the pseudomode expansion \eqref{pseudomode expansion} first appeared in the 1965 paper by Hazime Mori~\cite{Mori_1965_Continued-fraction}, exactly in the context of the recursion method. However, his approach to computing pseudomode eigenvalues and amplitudes (known as ``$n$-pole approximation''~\cite{viswanath2008recursion}) was based on assumptions that were, in fact, never satisfied in generic many-body systems. Later the pseudomode expansion  \eqref{pseudomode expansion}  recurrently reappeared   in various other contexts \cite{dalton2001theory, Meier_2012_Eigenmodes, Barocchi_2012_Exact,Barocchi_2014_Exponential,teretenkov2019pseudomode, pleasance2020generalized, Teretenkov_2024_Exact,Mori_2024_Liouvillian-gap}.


For a nonintegrable many-body system, one is able to accurately compute only a finite number of terms in eq.~\eqref{pseudomode expansion}. Fortunately, the pseudomode expansion typically converges very fast, as illustrated by examples below. As a result, the {\it truncated} pseudomode expansion containing only a small number $l=\lmax$ of the terms in the sum \eqref{pseudomode expansion} can deliver a very good approximation to the actual correlation function. Note that the asymptotic behavior of the autocorrelation function at $t\rightarrow \infty$ is, in general, determined by the leading eigenvalue $\lambda_0$ in case of $\im \lambda_0=0$, or by the leading pair of complex conjugate eigenvalues $\lambda_0=\lambda_1^*$ otherwise.

\medskip
\noindent{\it Toy model.}~ Here we consider a toy evolution superoperator \cite{Parker_2019,Balasubramanian_2022_Quantum,Bhattacharjee_2023_Operator,supp_pseudomode} with
\begin{equation}\label{toy Lanczos coefficients}
b_n=\sqrt{1-\gamma^2}\, n,\qquad  a_n=i \gamma (2n+1).
\end{equation}
While it is not obtained from any known microscopic many-body model, it is believed to grasp some of the major general features of actual evolution operators, since the sequences of  $b_n$ and $a_n$ follow the UOGH \cite{Parker_2019,bhattacharya2022operator,Liu_2023_Krylov,Bhattacharjee_2023_Operator}.

Note that coefficients $a_n$ in eq. \eqref{toy Lanczos coefficients} are not exactly of the form \eqref{a_n}. Most importantly, they do not vanish for any $n>0$. The consequence of this fact will be discussed below.

The set \eqref{Heisenberg equations} of Heisenberg equations for this toy $\L$ can be explicitly solved with the  result \cite{Parker_2019,Balasubramanian_2022_Quantum,Bhattacharjee_2023_Operator,supp_pseudomode}
\begin{equation}\label{C(t) toy model}
C(t)=\frac1{\cosh(t)\,(1+\gamma \tanh(t))}~\xrightarrow{\gamma\rightarrow 0}~ \frac1{\cosh(t)}.
\end{equation}
Our goal is to reproduce this autocorrelation function by means of the pseudomode expansion \eqref{pseudomode expansion}.

This task can be easily completed since  the spectrum and the eigenvectors of $\L$ can be explicitly computed for $\gamma\in(0,1)$ \cite{Parker_2019,Balasubramanian_2022_Quantum,Bhattacharjee_2023_Operator,supp_pseudomode}:
\begin{align}
\lambda_l & =-(2l+1), \label{eigenvalues toy model}\\
\U_l^0 &=
\sqrt{\frac{2}{1+\gamma}}\, (-i)^l\left(\frac{1-\gamma}{1+\gamma}\right)^{l/2} \,\xrightarrow{\gamma\rightarrow +0} \sqrt{2}\,(-i)^l.\label{eigenvectors toy model}
\end{align}
All eigenvalues are real negative and do not depend on $\gamma$, and $\U_l^0$ are  well-defined in the limit of vanishing dissipation strength. Hence our assumptions are satisfied and  the pseudomode expansion \eqref{pseudomode expansion}  reads
\begin{equation}\label{pseudomode expansion toy model}
C(t) =\frac1{\cosh(t)}=2\sum_{l=0}^\infty (-1)^l \, e^{-(2l+1)t}.
\end{equation}

The eigenvectors \eqref{eigenvectors toy model} exhibit localization for any finite $\gamma$ (see \cite{supp_pseudomode} for details) but not in the limit of vanishing $\gamma$. Still, the sum in eq. \eqref{pseudomode expansion toy model} converges for any fixed $t>0$, thanks to the linear growth of $\lambda_l$ with $l$. This is illustrated in Fig.~\ref{fig_toy_model}, where the truncated pseudomode expansion is compared to the actual value of $C(t)$ for various truncation orders $\lmax$. One can see that the convergence is spectacular at intermediate and large times. In particular, the leading pseudomode correctly reproduces the asymptotic behavior $C(t)\sim 2e^{-t}$ at large times. However, the convergence worsens at small times and is absent at $t=0$. This is not surprising, since the small-time behavior is expected to be reproduced provided  coefficients $a_n$  vanish for first few $n$, which is not the case in this toy model.

\

\medskip
\noindent{\it Spin models.}~ Here we consider two nearest-neighbour spin-$1/2$ models on the square lattice, the $XX$ model and  the  transverse field Ising model, with respective Hamiltonians given by
\begin{align}
H_{XX}& =\sum_{\langle \ii \jj \rangle} (\sigma^x_{\ii}\sigma^x_{\jj}+\sigma^y_{\ii}\sigma^y_{\jj}). \label{H 2D XX}\\
H_{\rm Ising}& =\sum_{\langle \ii \jj \rangle} \sigma^x_{\ii}\sigma^x_{\jj}+\sum_{\jj} \sigma^z_{\jj}, \label{H 2D Ising}
\end{align}
Here $\ii$ and $\jj$ label lattice sites and the sum over $\langle \ii \jj \rangle$ runs over bonds connecting neighbouring sites.

For both models we choose the  observable
\begin{equation}\label{observable}
A= \sum_{
\substack{~\,\langle \ii \jj \rangle_{-}\\
                  \ii \prec \, \jj
        }
}
(\sigma^x_{\ii} \sigma^y_{\jj}-\sigma^x_{\jj} \sigma^y_{\ii}),
\end{equation}
where the sum runs over horizontal bonds, the site $\ii$ being always to the left of the site $\jj$.
Up to multiplicative constants, this observable has the meaning of the energy current for the Ising model  and the spin current  for the $XX$ model.

To obtain the pseudomode expansion of the autocorrelation function, we proceed in several steps. First, we explicitly compute    $n_{\rm max}=11$  Lanczos coefficients $b_n$  for the $XX$  model \cite{supp_pseudomode}  and  $n_{\rm max}=21$ -- for the Ising model \cite{Uskov_Lychkovskiy_2024_Quantum,supp_pseudomode}.

Next, using $n_{\rm max}$ first Lanczos coefficients, we extrapolate the subsequent coefficients linearly, in accordance with the UOGH. Details of the extrapolation procedure and an estimate of the extrapolation error can be found in the Supplement~\cite{supp_pseudomode}.


Next we choose some finite $\gamma>0$ and thus fix the diagonal coefficients $a_n$ according to eq. \eqref{a_n}. Then we truncate the infinite matrix $\L$ at some  size $k$ (note that this truncation is unrelated to the truncation of the pseudomode expansion~\eqref{pseudomode expansion}) and compute its eigenvalues $\lambda_l(\gamma,k)$ and eigenvectors  $\U_l^n(\gamma,k)$ by numerical diagonalization. By repeating the latter step for larger and larger $k$, we numerically take the limit of $k \rightarrow \infty$ and obtain $\lambda_l(\gamma)=\lim\limits_{k\rightarrow \infty} \lambda_l(\gamma,k)$ and $\U_l^0(\gamma) =\lim\limits_{k\rightarrow \infty} \U_l^0(\gamma,k)$. The convergence is fast thanks to the localization:  essentially, the limit is saturated for $k \gtrsim  \kappa_l^{-1} + l$ \cite{supp_pseudomode}.

Finally, by repeating the above procedure for a range of $\gamma>0$, we numerically compute  eigenvalues $\lambda_l=\lim\limits_{\gamma\rightarrow+0} \lambda_l(\gamma)$ and amplitudes $\U_l^0=\lim\limits_{\gamma\rightarrow+0} \U_l^0(\gamma)$ for $l_{\rm max}$ first pseudomodes, as illustrated in  \ref{fig spin models} (b),(c),(f),(g). Plugging them in eq. \eqref{pseudomode expansion} and truncating the sum at  $l=l_{\rm max}$, we get the truncated pseudomode expansion of the autocorrelation function.

Remarkably, in contrast to the toy model, here the localization of eigenvectors survives the limit of vanishing coupling, and pseudomode amplitudes $\big(\U_l^0\big)^2$ decay exponentially with $l$, as can be seen from insets in Fig. \ref{fig spin models} (a),(e). This dramatically increases the rate of convergence of the pseudomode expansion. In fact, it turns out that  a correlation function can be accurately approximated by a sum of only a few first pseudomodes, as illustrated in Fig. \ref{fig spin models} (d),(h).

The latter fact is quite fortunate, since  the uncertainty in extrapolating the sequence of Lanczos coefficients limits the accuracy of the computed  pseudomodes eigenvalues $\lambda_l$ and amplitudes $\U_l^0$, and the corresponding error grows with  $l$, as illustrated in Fig \ref{fig spin models} (a),(e). As a result,  in practice only a finite number of first pseudomodes can be reliably recovered for a finite $n_{\rm max}$.  Naturally, this number grows when $n_{\rm max}$ is increased.


\medskip
\noindent{\it Summary.}~ To summarize, our studies confirm a long-standing surmise \cite{Mori_1965_Continued-fraction,Prosen_2002_Ruelle,Fine_2004_Long-time, Fine_2005_Long-time,Prosen_2007_Chaos,Barocchi_2012_Exact,Barocchi_2014_Exponential,Mori_2024_Liouvillian-gap} that a time-dependent correlation function of a generic  many-body system can be accurately approximated by a linear combination of a few pseudomodes $e^{\lambda_l t}$ with complex eigenvalues $\lambda_l$.
We have presented an algorithmic way to compute pseudomode eigenvalues and amplitudes starting from the Lanczos coefficients obtained within the recursion method.   We have applied the method to two quantum spin models on a square lattice. In these examples, the knowledge of a dozen or two of first Lanczos coefficients is enough to find a few first pseudomodes  and compute the autocorrelation function with a good accuracy.





\medskip
\noindent{\it Note added.}~ After the first version of the present manuscript had been posted to arXiv and submitted to the journal,  five papers \cite{dodelson2024ringdown,Znidaric_2024_Momentum-dependent,Jacoby_2025_Spectral,zhang2024thermalization,yoshimura2025theory} pursuing a closely related agenda appeared. Various systems were addressed there, including the Sachdev-Ye-Kitaev model \cite{dodelson2024ringdown}, one-dimensional kicked Ising model \cite{Znidaric_2024_Momentum-dependent} and random unitary circuits \cite{Jacoby_2025_Spectral,zhang2024thermalization,yoshimura2025theory}.


\medskip
\begin{acknowledgments}
\noindent{\it  Acknowledgments.}
We thank Boris Fine, Nikolay Il'in,  Alexander Jacoby and Matthew Dodelson for valuable discussions. This work was supported by the Russian Science Foundation under grant \textnumero~24-22-00331, \url{https://rscf.ru/en/project/24-22-00331/}
\end{acknowledgments}

\bibliography{C:/D/Work/QM/Bibs/open_systems,C:/D/Work/QM/Bibs/recursion_method}

\clearpage

\onecolumngrid

\renewcommand{\theequation}{S\arabic{equation}}
\stepcounter{myequation}

\renewcommand{\thefigure}{S\arabic{figure}}
\stepcounter{myfigure}

\renewcommand{\thetable}{S\arabic{table}}
\setcounter{table}{0}

\setcounter{page}{1}

\appendix

\section{
	{\large Supplementary material}
	\\
	to
	``Pseudomode expansion of many-body correlation functions''\\
	  by Alexander Teretenkov, Filipp Uskov and Oleg Lychkovskiy
}


\section{S1. Toy model}
The toy evolution operator \eqref{L},\eqref{toy Lanczos coefficients} can be represented as
\begin{align}
		\L = i \gamma M + \sqrt{1 - \gamma^2} (B + B^{\dagger}),
\end{align}
where
\begin{equation}\label{M and B}
M=
\begin{pmatrix}
    1  & 0 & 0  & 0  & \dots  \\[0.3 em]
   0 & 3  & 0 & 0  & \dots \\[0.3 em]
     0 & 0  & 5 &  0  &  \dots \\[0.3 em]
     0 & 0  & 0 & 7  &  \dots \\[0.3 em]
\vdots & \vdots & \vdots & \vdots & \ddots
  \end{pmatrix},
\qquad
B=
\begin{pmatrix}
     0  & 1 & 0 & 0  & \dots  \\[0.3 em]
     0 & 0  & 2 & 0  & \dots \\[0.3 em]
     0 & 0  & 0 &3  &  \dots \\[0.3 em]
     0 & 0  & 0 & 0  &  \dots \\[0.3 em]
\vdots & \vdots & \vdots & \vdots & \ddots
  \end{pmatrix}.
\end{equation}
Crucially, operators $M$ and $B$ satisfy
\begin{equation}\label{commutation toy model}
		[B, B^{\dagger}] = M, \qquad [B, M] =  2 B , \qquad [B^{\dagger}, M] = - 2 B^{\dagger},
\end{equation}
{\it cf.} ref. \cite{accardi2002renormalized}.
These commutation relations can be easily obtained from the action of the above operators on basis vectors $|n\rangle$, $n=0,1,2,\dots$:
\begin{equation}
		B |n\rangle =  n |n-1\rangle,\quad B^{\dagger} |n\rangle =  (n +1) |n+1\rangle,\quad M  |n\rangle = (2n+1) |n\rangle.
\end{equation}

Using  commutation relations \eqref{commutation toy model}, one brings  $\L$ to the diagonal form as follows:
\begin{equation}
\L	=  e^{ i \tau(\gamma) (B^{\dagger} - B)} (i M)   e^{ - i\tau(\gamma)  (B^{\dagger}  - B)},
\end{equation}
where
\begin{equation}
	\tau(\gamma) = \arccos \left(- \sqrt{\frac{1 +\gamma}{2}} \right),\quad \gamma\in(0,1).
\end{equation}
Thus $\U$ and $ \Lambda$ from  eq. \eqref{L in diagonal form} read
\begin{equation}
\U=e^{ i \tau(\gamma) (B^{\dagger} - B)},\qquad \Lambda=-M.
\end{equation}
Eigenvalues \eqref{eigenvalues toy model} follow immediately. Eigenvectors \eqref{eigenvectors toy model} and autocorrelation function \eqref{C(t) toy model} are obtained after some straightforward algebra.

\section{S2. Pseudomodes in  the $XX$ and Ising models on the square lattice}
\subsection{S2.1. Computing $n_{\rm max}$ first Lanczos coefficients}

We use a computer program reported in  ref. \cite{Uskov_Lychkovskiy_2024_Quantum} to compute $n_{\rm max}$ first moments and Lanczos coefficient. We are able to achieve $n_{\rm max}=11$ for the  $XX$ model and $n_{\rm max}=21$ for the Ising model.  The results are shown in Table~\ref{table XX} and Table~\ref{table Ising}, respectively. The Lanczos coefficients are also shown in Fig. \ref{fig fitting}(a),(d).

\clearpage

\begin{table}[t]
\begin{tabular}{l|l|l}
    {$n$}~~ & {$\mu_{2n}$} & $b_n$ \\
    \hline
    1  & 16 & 4.0\\
    2  &  1024 &  6.9282\\
    3  &  136192 & 9.5917\\
    4  &   30277632    &  12.2131\\
    5  &   9983098880 & 14.5925\\
    6  &   4498943967232 & 16.6669\\
    7  &     2599858319392768 & 18.1310\\
    8  &      1834298158063026176 & 19.9233\\
    9  &  1527776305296896425984 & 22.1031\\
    10 &     1472244134830457072123904 & 24.0962\\
    11 &   1624011303835553036301762560~~ & 25.8961\\
\end{tabular}
\caption{Moments and Lanczos coefficients for the $XX$ model on the square lattice, eq. \eqref{H 2D XX}, and observable \eqref{observable}.\label{table XX}}
\end{table}

\bigskip

\begin{table}[h]
\begin{tabular}{l|l|l}
    {$n$}~~ & {$\mu_{2n}$} & $b_n$ \\
    \hline
    1  & 8 & 2.8284\\
    2  &  192 &  4.0\\
    3  &  8192 &  5.2915\\
    4  &    546816   & 6.7612  \\
    5  &   53321728 & 8.0861\\
    6  &  7175143424  & 9.3673\\
    7  &   1268516651008   &  10.6283 \\
    8  &    282984740552704  & 11.7194 \\
    9  & 77099370874404864  &12.8275 \\
    10 &    24994506356392198144  & 13.9370\\
    11 &  9447181150949853888512 &  14.9831\\
    12  & 4097682751836764564881408  &   15.8407\\
    13  & 2013581261449707414815244288 & 16.6125\\
    14  & 1108435050944563799272751890432     &  17.3425\\
    15  & 676413174916563342565009178755072  & 18.2729\\
    16  & 453089374036703756783428077313064960  & 19.1098\\
    17  & 330151040493164361120913228230360563712   & 19.9837\\
    18  & 259674454740125766898519426917750737469440    &  20.7966\\
    19  & 219085451111188700544806075134247988359069696 &  21.6735\\
    20 &  197334932443983424783725995684533384746838786048   & 22.5520\\
    21 & 189115452873144653249584832927971057967336192475136 & 23.4344\\
\end{tabular}
\caption{Moments and Lanczos coefficients for the Ising model on the square lattice, eq. \eqref{H 2D Ising}, and observable \eqref{observable}.\label{table Ising}}
\end{table}

\clearpage

\subsection{S2.2. Extrapolating further Lanczos coefficients according to the UOGH}

\begin{figure*}[t] 
		\centering
		\includegraphics[width=0.25 \linewidth]{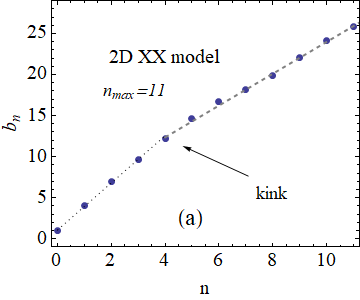}
~~
		\includegraphics[width=0.3 \linewidth]{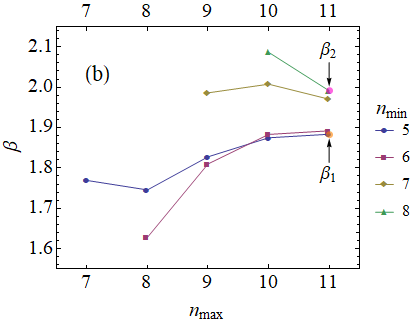}
~~
        \includegraphics[width=0.3 \linewidth]{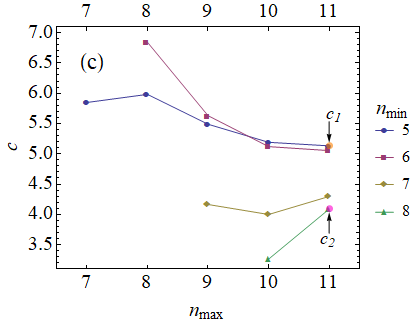}\\[1em]
		\includegraphics[width=0.25 \linewidth]{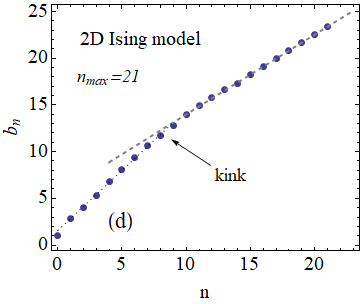}
~~
		\includegraphics[width=0.3 \linewidth]{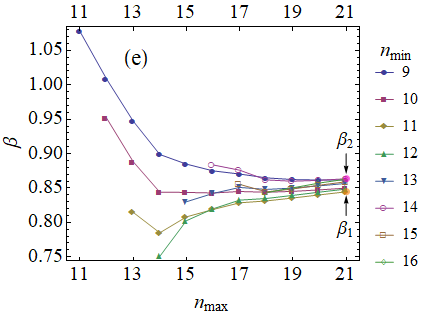}
~~
        \includegraphics[width=0.3 \linewidth]{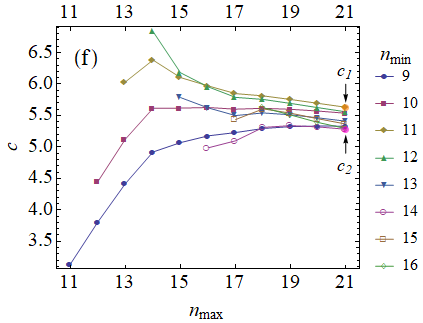}
		\caption{Lanczos coefficients and linear fits to these coefficients for the $XX$ (upper row) and the Ising (lower row) models on the square lattice. (a),(d): $n_{\rm max}$ Lanczos coefficients $b_n$ are shown by blue points, linear fit to these coefficients is shown with a gray dashed line. One can see a kink separating two regions with different slopes. Only points to the right of the kink are used for the fit. (b),(c),(e),(f): Dependence of fitting parameters on the maximal number $n_{\rm max}$ of the available moments, for various fitting thresholds $n_{\rm min}$. Fitting parameters $(\beta_1,c_1)$ and $(\beta_2,c_2)$ used to estimate the uncertainty due to fitting errors are shown, respectively, by large orange and pink points and marked by arrows.}
		\label{fig fitting}
\end{figure*}

We extrapolate Lanczos coefficients $b_n$ for $n>n_{\rm max}$ according to the UOGH, i.e. linearly:
\begin{equation}\label{fit}
b_n^{\rm extr}=\beta\, n+c,\qquad n\geq n_{\rm max}+1.
\end{equation}

Constants $\beta$ and $c$ are obtained from the least squares fit of $b_n$ for $n_{\rm min}\leq n \leq n_{\rm max}$. The choice of $n_{\rm min}$ is somewhat intricate. Examining the  plots of Lanczos coefficients for the two models under consideration, Fig. \ref{fig fitting}(a),(d), we see that each of them exhibits a kink at some $n=n_{\rm kink}$. The kink separates two regions of $n$ with apparently different slopes $\beta$. The origin of this phenomenon is an interesting open question. We assume that there are no other kinks for $n>n_{\rm max}$, and thus the Lanczos coefficient for $n_{\rm kink}\leq n \leq n_{\rm max}$ can be used for the fit.

With a finite number of explicitly computed Lanczos coefficients available, the values of $\beta$ and $c$ extracted from the fit are necessarily somewhat off the true asymptotic values. To access the uncertainty of fitted $\beta$ and $c$, we perform fits for various $n_{\rm min}\geq n_{\rm kink}$ and $n_{\rm max}$, with the results shown in Fig. \ref{fig fitting}(b),(c),(e),(f). For a given $n_{\rm max}$, we estimate the uncertainty in determining $\beta$ and $c$ as the maximal spread of these values for various $n_{\rm kink}\leq n_{\rm min}\leq n_{\rm max} - 2 $. This uncertainty tends to decrease with increasing $n_{\rm max}$, as can be seen from Fig. \ref{fig fitting}(b),(c),(e),(f).

\begin{figure*}[t] 
		\centering
		\includegraphics[width=0.45 \linewidth]{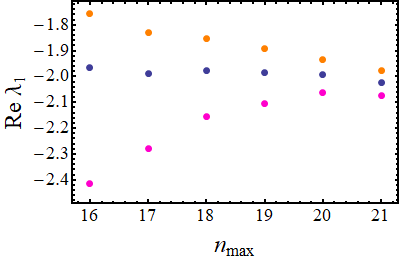}
~~~~
		\includegraphics[width=0.45 \linewidth]{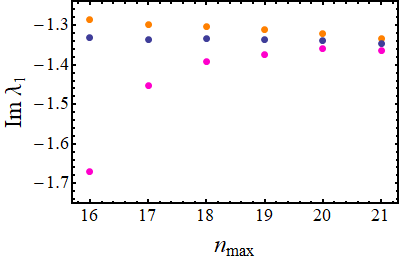}
		\caption{Extrapolation uncertainty of a pseudomode eigenvalue as a function of the number $n_{\rm max}$  of explicit Lanczos coefficients available. Shown are real (left) and imaginary (right) parts of the first pseudomode eigenvalue $\lambda_1$ for the Ising model on the square lattice. Orange, pink and blue points correspond to fitting parameters $(\beta_1,c_1)$, $(\beta_2,c_2)$ and $(\beta_{\rm av},c_{\rm av})$, respectively. This color code is consistent with that of Fig. \ref{fig fitting} and Fig. \ref{fig spin models}. }
		\label{fig fit error}
\end{figure*}

To access the effect of extrapolation uncertainty, we compute pseudomodes and autocorrelation functions for three sets of fitting parameters:
$(\beta_1,c_1)$, $(\beta_2,c_2)$ and $\big(\beta_{\rm av}=(\beta_1+\beta_2)/2,c_{\rm av}=(c_1+c_2)/2\big)$. The first two fits are extremes obtained by varying $n_{\rm min}$, see Fig. \ref{fig fitting} (b),(c),(e),(f). The third fit is  the average of the first two ones. The effect of extrapolation uncertainty on pseudomode eigenvalues is shown in Fig. \ref{fig spin models}(a),(e) (see also Fig. \ref{fig fit error}), while the effect on correlation functions -- in  Fig. \ref{fig spin models}(d),(h) . Whenever not specified explicitly, we use the average fit $\big(\beta_{\rm av},c_{\rm av})$.

Naturally, the more Lanczos coefficients can be computed explicitly, the less uncertain the fit is and more accurate  pseudomode eigenvalues and amplitudes are. We illustrate the accuracy of the pseudomode eigenvalue restoration in Fig. \ref{fig fit error}.


\subsection{S2.3. Taking thermodynamic limit numerically}

\begin{figure*}[t] 
		\centering
		\includegraphics[width=0.4 \linewidth]{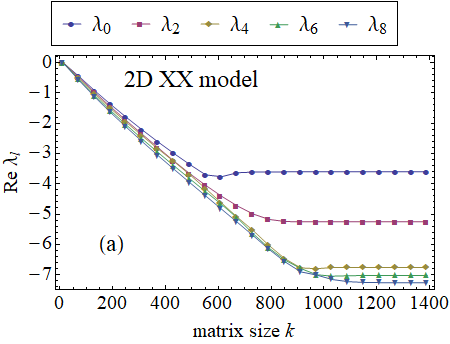}
~~
		\includegraphics[width=0.4 \linewidth]{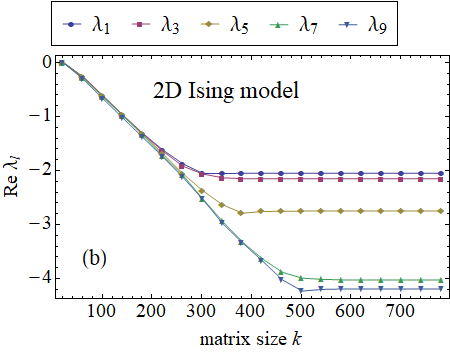}\\[1em]
		\includegraphics[width=0.4 \linewidth]{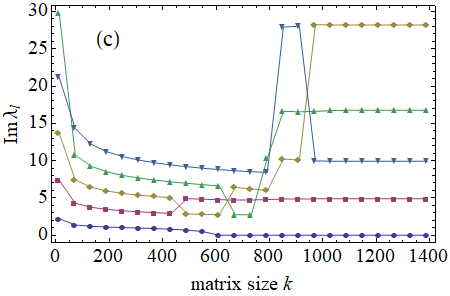}
~~
		\includegraphics[width=0.4 \linewidth]{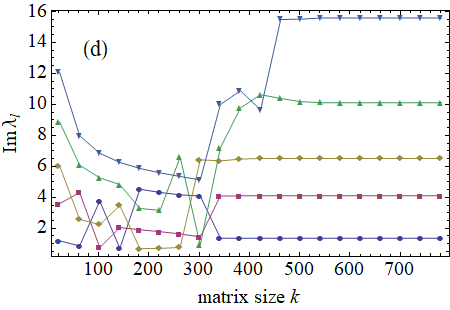}\\[1em]
		\includegraphics[width=0.4 \linewidth]{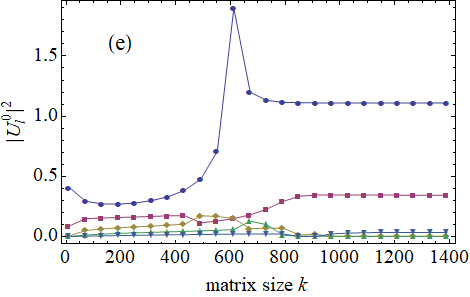}
~~
		\includegraphics[width=0.4 \linewidth]{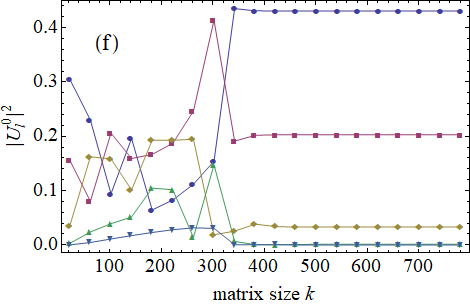}\\[1em]
		~~~~~\includegraphics[width=0.38 \linewidth]{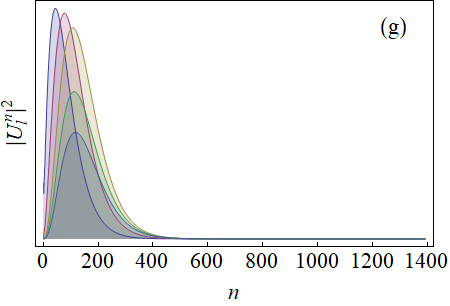}
~~~~~
		\includegraphics[width=0.38 \linewidth]{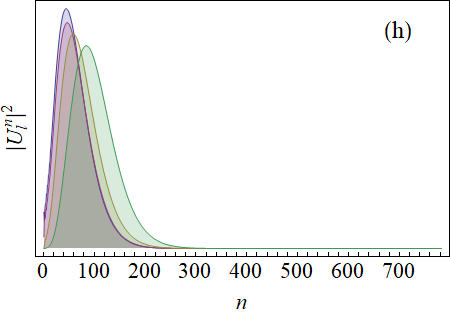}\\[1em]
		\caption{The numerical thermodynamic limit for pseudomode eigenvalues and amplitudes for a fixed $\gamma=0.05$. Left (right) column -- $XX$ (Ising) model on the square lattice.  First two rows show the scaling of, respectively, the real and the imaginary  parts of the first few pseudomode eigenvalues $\lambda_l(\gamma,k)$ with the size $k$ of the truncated matrix $i\cal L$. The third row shows the scaling of the absolute value of the pseudomode amplitudes $\left|{{\cal U}_l^0(\gamma,k)}\right|^2$. The fourth row shows the respective full eigenvectors $\left|{{\cal U}_l^n}(\gamma,800)\right|^2$ as functions of $n$ for a fixed truncation size $k=800$ (the absolute scale is arbitrary for each eigenvector; the color code is consistent with that in other plots). One can see that, thanks to the eigenvector localization, the  pseudomode eigenvalues and amplitudes in (a)-(f) are abruptly saturated for $k$ exceeding the support of the respective eigenvectors in (g),(e).   }
		\label{fig size scaling}
\end{figure*}

In Fig. \ref{fig size scaling} we illustrate our procedure to numerically take the thermodynamic limit. One can see that, for a given $\gamma$, the values of $\lambda_l(\gamma, k)$ and $U^0_l(\gamma, k)$ abruptly saturate above some threshold size of the truncated matrix $i{\cal L}$. As can be seen from Fig. \ref{fig size scaling}, this threshold size is in fact determined by the support of the corresponding eigenvectors which is finite thanks to the localization.

\begin{figure*}[t] 
		\centering
		\includegraphics[width=0.32 \linewidth]{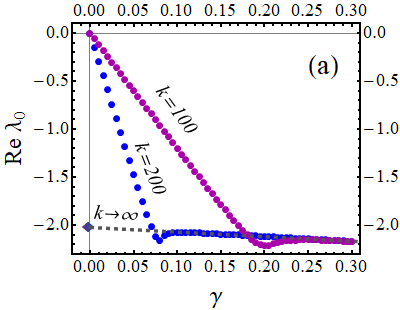}
~
		\includegraphics[width=0.32 \linewidth]{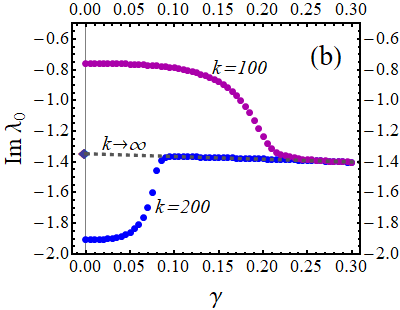}
~
        \includegraphics[width=0.3 \linewidth]{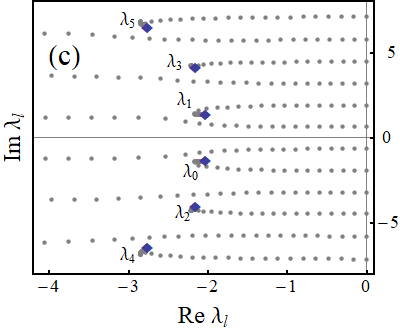}
		\caption{Dependence of pseudomode eigenvalues on $\gamma$ for a fixed size $k$ of the truncated matrix $i{\cal L}$ in the Ising model. (a) and (b): dependence of the real (a) and imaginary (b) parts of $\lambda_0$ on $\gamma$ for the fixed sizes $k=200$ (blue points) and $k=100$ (magenta points). The fit to $k\rightarrow\infty$ is shown by dashed line (see Fig. \ref{fig spin models}(f),(g)). (c) The flow of eigenvalues with $\gamma$ for $k=200$. The value of $\gamma$ changes from $0$ to $0.3$, consistent with (a),(b).  At zero $\gamma$ all eigenvalues are purely imaginary. For nonzero $\gamma$ eigenvalues acquire a negative real part. With the increase of $\gamma$,  eigenvalues tend to flow to the left.  In all three plots blue diamonds mark the ``correct'' limit $\lim\limits_{\gamma\rightarrow 0} \lim\limits_{k\rightarrow \infty} \lambda_l$ (see Fig. \ref{fig spin models}(f),(g)). Note that while a half of eigenvalues flow towards the rightmost end of the complex spectrum which we are interested in, the other half  quickly flows towards the leftmost part of the spectrum.  }
		\label{fig wrong order}
\end{figure*}

One may wonder what happens if one takes an ``incorrect'' order of limits:  $ \lim\limits_{k\rightarrow \infty} \lim\limits_{\gamma\rightarrow 0}\lambda_l$ instead of $\lim\limits_{\gamma\rightarrow 0} \lim\limits_{k\rightarrow \infty} \lambda_l$. Naturally, in this case eigenvalues would lose their negative real parts and become purely imaginary. Some amazing features of this limit are  illustrated in Fig. \ref{fig wrong order}.

\section{S2.4. Comparison between pseudomode expansion and extrapolated recursion method }

\begin{figure*}[t] 
		\centering
		\includegraphics[width=0.45 \linewidth]{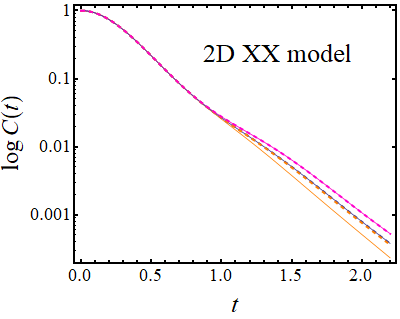}
~~~~
		\includegraphics[width=0.45 \linewidth]{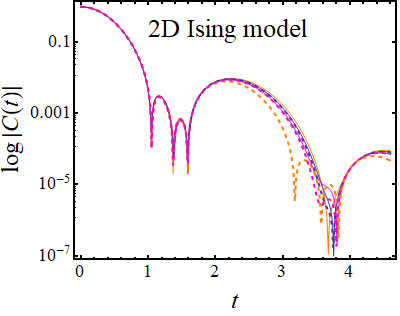}
		\caption{Correlation function computed by means of the truncated pseudomode expansion (solid) and  the extrapolated recursion method \cite{Uskov_Lychkovskiy_2024_Quantum} with $\cal L$ truncated at the size $k=1500$ (dashed). Orange, pink and blue lines correspond to fitting parameters $(\beta_1,c_1)$, $(\beta_2,c_2)$ and $(\beta_{\rm av},c_{\rm av})$, respectively. This color code is consistent with that of Fig. \ref{fig fitting} and Fig. \ref{fig spin models}. }
		\label{fig correlation functions}
\end{figure*}

One can compute the autocorrelation function by extrapolating Lanczos coefficients according to eq. \eqref{fit}, truncating the system of  coupled Heisenberg equations \eqref{Heisenberg equations} at high $k\sim 1000$ and numerically solving the truncated system \cite{Uskov_Lychkovskiy_2024_Quantum}. We compare thus computed autocorrelation function with the pseudomode expansion \eqref{pseudomode expansion} in Fig.\ref{fig correlation functions}. One can see that the agreement is good, with the discrepancy between the two approaches not exceeding the uncertainty due to the extrapolation of the Lanczos sequence. We conclude that the two approaches to computing the autocorrelation function have a comparable accuracy.

\end{document}